\documentclass{article}

\newtheorem{prop}{Proposition}

\def\C{{ \! \rm \ I\!\!\!C}}

\def\R{{ \! \rm \ I\!R}}

\def\l{{ \lambda}}
\def \square{\hbox {$\sqcup $\llap {$\sqcap $}}}
\def \endsquare{\ifmmode \kern 0.5em\square \else
                \discretionary{}{\hbox to\hsize {\hfill\square}\null}
                                {\kern 0.5em\square}\fi }

\title{{Some Remarks about Duality, Analytic Torsion and Gaussian
Integration in Antisymmetric Field Theories}} 
\author{Alexander Cardona}

\date{\small{\it  Laboratoire de Math\'ematiques Appliqu\'ees
\\Universit\'e Blaise Pascal (Clermont II) 
\\ Complexe Universitaire des C\'ezeaux \\ 63177 Aubi\`ere Cedex, France.\\
{\rm cardona@ucfma.univ-bpclermont.fr}}}
\begin{document}

\maketitle
\begin{abstract}

From a path integral point of view (e.g. \cite{Q98}) physicists have 
shown how {\it duality} in antisymmetric quantum field theories on a closed
space-time manifold $M$ relies in a fundamental
way on Fourier Transformations of formal infinite-dimensional volume 
measures. We first review these facts from a measure theoretical point of 
view, setting the importance of the Hodge decomposition theorem in the 
underlying geometric picture, ignoring the local symmetry which lead to 
degeneracies of the action. To handle these degeneracies we then apply 
Schwarz's Ansatz showing how duality leads to a
factorization of the analytic torsion of
$M$ in terms of the partition functions associated to degenerate ``dual" 
actions, which in the even dimensional case corresponds to the 
identification of these partition functions.
\end{abstract}

\maketitle

\section*{Introduction}

Antisymmetric field theories are generalizations of electromagnetic
theory where the potential $1$-form is replaced by
a $k$-form. Some remarkable facts arising in electromagnetism
are also observed in general antisymmetric theories, notably T-{\it
duality} on which we will focus here. In electromagnetic theory this
type of duality  corresponds to the observation that electric and
magnetic fields in the  theory are interchanged under transformations
taking solutions of field equations into solutions of the Bianchi identity,
particles into topological defects, weak couplings into strong
couplings, etc. (for a review see \cite{O95}). \\ \\ 
Consider a theory
of antisymmetric tensors on a $n$-dimensional space-time manifold $M$
equpped with a Riemannian metric.  Let
$\omega_{i_1i_2 \dots i_{k}}$ be a $k$-tensor field on $M$, consider
the $k$-form 
$$ \omega_{k} = \omega_{i_1i_2 \dots
i_{k}} dx^{i_1} \wedge dx^{i_2} \wedge \cdots \wedge dx^{i_{k}},$$ 
for $0 \le k \le n$, and the {\it Euclidean Action} of the theory
defined by
\begin{equation}
{\cal S}(\omega_k) = \langle d_k \omega_k, d_k \omega_k \rangle,
\end{equation}
where $d_k$ denotes the exterior derivative on the space $\Omega^k$ of
$k$-forms on $M$ and the inner product
$( \, ,) : \Omega^k \times \Omega^k \rightarrow \R$ is 
defined by Hodge-star operation on $ \Omega^k$, namely
\begin{equation}
 ( \alpha_k, \beta_k ) = \int_M \alpha_k \wedge *
\beta_k  . 
\end{equation}
This generalizes electromagnetic theory, where the 
potential is described by a $1$-form $A$, the electromagnetic field by
its exterior derivative ($ F= dA$), and
where by {\it gauge invariance} of the theory we mean the invariance of
$F$ under ``gauge" transformations on $A$ of the form
\begin{equation} 
A \mapsto A + d \chi ,
\end{equation}
$\chi$ being an arbitrary function ($0$-form) on $M$.
\\ \\
Following \cite{Q98}, T-duality in the case of antisymmetric field
theories is the statement that two different theories (defined by two
different actions ${\cal S} $ and $ {\cal S}^* $) give rise to
the same generating function, being therefore (at the quantum level)
physically equivalent. As in \cite{Q98} \cite{W99} and many other 
references on this topic, in this paper we focus on the
identification of  partition functions, hoping to complete the
discussion on the level of generating functions in some later work. 
\\ \\
By partition function we understand the formal  object
\begin{equation}
Z(S) = \int \exp \left\{- {\rm k} \,{\cal S} (\alpha)\right\} \, 
[{\cal D}\alpha] 
\end{equation}
where ${\rm k}$ denotes a
(positive) constant (including Planck's and coupling constants), 
$[{\cal D}\alpha]$ denotes a formal measure on the space of all the
fields $\alpha$ and $\cal S (\alpha )$ the classical action of the
theory under consideration.  Looking for a dual version of a theory
means looking for a different action, called {\it dual} action (on a 
different set of {\it dual} fields), giving rise to a {\it dual}
partition function. Starting from a given action (i.e a given theory),
a standard procedure to obtain a dual action (i.e. a dual theory) is
the so-called {\it gauging} of the global symmetry of the original
theory \cite{Q98}\cite{W99}. This requires introducing new variables
into the original action in such a way that integrating them out we
can recover the original theory and integrating out the original
variables of the action we find the dual one. Unlike in \cite{Q98}, we 
only consider local symmetries (namely of the type $(3)$) of the classical 
action, having left aside global symmetries because of the acyclicity 
assumption (see section $1$). The very presence of local symmetries leads 
to degenerate actions. Forgetting about degeneracy of the classical action, 
as was pointed out in \cite{Q98}, duality strongly relies on Fourier 
transformations of measures. We follow this point of view in section $2$. 
On the other hand, if one wants to take into account the presence of local 
symmetries, a method is required to handle partition functions with 
degenerate actions. In the context of what is now called Topological 
Quantum Field Theories, Schwarz proposed an Ansatz to compute such partition 
functions which we apply in section $3$, this leading us to an interpretation 
of duality of partition functions in terms of a factorization of the analytic 
torsion of the underlying space-time manifold. In even dimensions this give 
the expected identification of the partition function of an action with its 
dual.
\\ \\
Let us describe briefly the contents of this contribution. In section
$1$ we describe the geometric setting underlying the definition
of antisymmetric tensor fields. In section $2$ we give a measure
theoretical interpretation of the formal path integral manipulations
in the case of duality between two antisymmetric field theories
defined by non degenerate action functionals and, following Quevedo
\cite{Q98}, we give the heuristic path integral interpretation of
duality in terms of Fourier transformation of measures. In section $3$ we 
use the approach proposed by Schwarz
\cite{S79} to study the partition function of a degenerate
functional, and we show how two dual actions yield a factorisation of
the analytic torsion on the underlying manifold. \\
To distinguish between
formal  (heuristic) equalities from precise mathematical ones we shall
use the symbol 
$``="$ for the first kind.

\section{The Geometric Setting} 
Consider a closed (i.e. compact and without
boundary) $n$-dimensional Riemannian manifold
$M$, and let $\rho$ be a representation of the fundamental group of
$M$ on an inner product vector space $V$.
Let $ E(\rho)$ be the vector bundle over $M$ defined by $\rho$, and
consider the space of $k$-forms on $M$ with values in $E(\rho)$, for
$0 \le k \le n$, i.e.
$C^\infty$-sections of the vector bundle $\Lambda^{k} T^*M \otimes 
E(\rho) $. This space of sections, that we will denote by
$\Omega^k$, will be the space of $k$-antisymmetric tensor fields. The
bundle $E(\rho)$ comes with a flat connection that couples with
exterior differentiation on $k$-forms to define an exterior
differential (also denoted $d_k$), on 
$E(\rho)$-valued $k$-forms, such that $d_k^2 = 0$. The inner
product $(2)$ on $\Lambda^{k} T^*M $, defined by the Riemannian metric
on $M$, thogether with the inner product on $E(\rho)$, provides
$\Omega^k$ with an inner product that we will denote by $\langle \, ,
\rangle$. With respect to that inner product, and by Hodge
$*$-duality map, $d_k^* = (-1)^{nk + n+1}*d_{n-k-1}*$ defines the
formal adjoint of $d_k$. Finally, we assume that the 
complex
\begin{equation}
\begin{array}{cccccccccccccccc}
 0  \!\!\! & \! 
\stackrel{}{\longrightarrow} \!\!\! & \! \Omega^0 \!\!\! & \!
\stackrel{d_0}{\longrightarrow} \!\!\! & \! 
\cdots \!\!\! & \! \Omega^{k-1} \!\!\! & \!
\stackrel{d_{k-1}}{\longrightarrow} \!\!\! & \! \Omega^{k} \!\!\! & \! 
\stackrel{d_{k}}{\longrightarrow} \!\!\! & \! \Omega^{k+1} \!\!\! & \!
\stackrel{d_{k+1}}{\longrightarrow} \!\!\! & \! 
\cdots \!\!\! & \Omega^{n}\!\!\! & \!
\stackrel{d_{n}}{\longrightarrow} \!\!\! & \! 0 ,
\end{array} 
\end{equation}
is acyclic, i.e. all the de
Rham cohomology groups of the complex are trivial ($H^k(M, \rho)=
\{0\}$, $0 \le k \le n $). This representation of $\pi_1(M)$ will be
fixed through all  the paper and no specific
reference to it will be given (in the notation) in the sequel. 
\\ \\
Let us focus on the space of $k$-forms
$$\Omega^{k-1} \stackrel{d_{k-1}}{\longrightarrow}  \Omega^{k}  
\stackrel{d_{k}^*}{\longleftarrow}  \Omega^{k+1}, $$
where $d^*_k$ denotes the formal adjoint to $ d_k$, and on the Hodge
decomposition
\begin{equation}
 \Omega^k = \Omega_{k}'  \oplus \Omega_{k}'' 
\end{equation}
where $ \Omega_{k}' = {\rm Im}\ d_{k-1}= {\rm Ker}\  d_k$
and $ \Omega_{k}''= {\rm Im}\ d_{k}^* = {\rm Ker}\  d_{k-1}^*$, as
follows from our asumption of acyclicity. Accordingly, 
$\omega_k \in \Omega^k$ splits into $ \omega_k =
\omega_{k}' \oplus \omega_{k}''$ where  $\omega_{k}' = d_{k-1}
\omega_{k-1} \in \Omega_{k}'$ and $ \omega_{k}'' = d_{k}^*
\omega_{k+1} \in \Omega_{k}''$, for some $\omega_{k-1} \in
\Omega^{k-1}$, $\omega_{k+1} \in \Omega^{k+1}$. \\ \\
Consider the functional
\begin{eqnarray} 
{\cal S}_o: \Omega^k & \rightarrow & \; \R \nonumber \\
 \omega_k & \mapsto & {\cal S}_o(\omega_k) = \langle \omega_k ,
\omega_k
\rangle ,
\end{eqnarray} 
on $k$-antisymmetric tensor fields. Then, from the 
decomposition $(6)$ and $d_{k} d_{k-1} =  d^*_{k-1}d^*_{k} = 0$, it
follows that
\begin{eqnarray*} 
 {\cal S}_o(\omega_k) & = & \langle d_{k-1} \omega_{k-1} \oplus
d_{k}^*
\omega_{k+1}, \; d_{k-1} \omega_{k-1} \oplus d_{k}^* \omega_{k+1}
\rangle\\  
             & = & \langle d_{k-1} \omega_{k-1} , \, d_{k-1}
\omega_{k-1} \rangle \oplus \langle d_{k}^* \omega_{k+1}, \, d_{k}^*
\omega_{k+1}\rangle.
\end{eqnarray*}
Thus, we find a canonical decomposition of $ \cal S$ in terms of two
degenerate action functionals, namely  
\begin{equation}
 {\cal S}_o(\omega_k)={\cal S}(\omega_{k-1}) \oplus
{\cal S}^*(\omega_{k+1}), 
\end{equation}
where
\begin{equation}
{\cal S}(\omega_{k-1})  = \langle d_{k-1} \omega_{k-1}
, \, d_{k-1} \omega_{k-1} \rangle 
\end{equation}
and
\begin{equation}
{\cal S}^*(\omega_{k+1}) = \langle d_{k}^*
\omega_{k+1}, \, d_{k}^* \omega_{k+1} \rangle, \;\;\;\;\;\;\; 
\end{equation}
which are degenerate on $\Omega^{k-1}$ and $\Omega^{k+1}$,
respectively. The functionals ${\cal S}(\omega_{k-1})$ and  $ {\cal
S}^*(\omega_{k+1})$ are degenerate but, by restriction on the
respective domains, the maps 
\begin{equation}
 d_k : \Omega_{k} '' \rightarrow \Omega_{k+1}'
\end{equation}  
and 
\begin{equation}
 d_k^* : \Omega_{k+1} ' \rightarrow \Omega_{k}'' ,
\end{equation}
are isomorphisms, giving rise to the bijective maps
$$d_{k-1}^*d_{k-1} :\Omega_{k-1} '' \; \rightarrow \;
\Omega_{k-1}'',  $$
$$ \;\;\;\;\; d_k d_k^* : \Omega_{k+1} ' \rightarrow
\Omega_{k+1}'.$$  
Thus, the functionals
\begin{equation}
 \widehat{{\cal S}}(\omega_{k-1}'')  = \langle d_{k-1}
\omega_{k-1}'' , \, d_{k-1} \omega_{k-1}'' \rangle 
\end{equation}
and
\begin{equation}
 \widehat{\cal S}^*(\omega_{k+1}') = \langle d_{k}^*
\omega_{k+1}', \, d_{k}^* \omega_{k+1}' \rangle, \;\;\;\;\;\;\; 
\end{equation}
are non-degenerate on ${\Omega_{k-1}''}$ and
$\Omega_{k+1}'$, respectively. These spaces are the ones we shall be 
working with in section $2$ in order to have partition functions of 
non-degenerate actions.\\
\\
The identification between two dual antisymmetric field theories involves 
identifying formal 
integrals, which we will interpret as gaussian integrals since they are 
defined using quadratic actions. In section $2$ we study the ``equivalence''
between two such partition functions from a measure theoretical point
of view in the case in which the action functionals involved are not
degenerate and, following Quevedo \cite{Q98}, we give the heuristic
path integral interpretation of duality. The case of degenerate
action functionals will be studied in section $3$.

\section{Duality and Gaussian Measures on Antisymmetric
Tensor  Fields}
\subsection{Some Facts about Gaussian Measures} 
A {\it characteristic function} on a topological vector space $E$ is a
continuous (on every finite dimensional subspace of $E$) function
$\chi$ satisfying
$$ \sum_{j,k=1}^N \alpha_j \bar{\alpha_k} \, \chi (\xi_j - \xi_k ) \geq 0$$ 
for $\alpha_k \in \C$, $ \xi_j \in E \;$  ($ j, k = 1 ,..., N$). In a finite 
dimensional vector space $E$, with inner product $\langle
\, , \rangle$, {\it Bochner's theorem} assures a one-to-one
correspondence between characteristic functions and
measures
\cite{Y85}. In particular, to the function
$$\chi (\xi) = \exp \left\{-{1 \over 2} \langle \xi , \xi 
\rangle \right\} $$
there corresponds a unique Borel measure on $E$, called {\it Gaussian
Measure} and denoted by  $ \mu$, such that
$$ \chi (\xi) = \int_{E} \; \exp \left\{i \langle \xi,
\phi \rangle \right\} d \mu (\phi) $$
and $ \mu (E) = 1$. In infinite dimensions, starting from a
characteristic function $\chi$ on  a topological vector space $E$, one
typically ends up with a measure with support in a larger space.  Even
in the case of a Hilbert space ${\cal H}$, the corresponding measure to a
characteristic function lies in some Hilbert-Schmidt extension of ${\cal H}$.
Bochner's theorem holds in the case of continuous
characteristic functions on a nuclear Hilbert space (a topological vector 
space whose topology is defined
by a family $ \{ \vert \vert \, \cdot \,\vert \vert_\alpha\}$ of Hilbertian 
semi-norms such that $ \forall \alpha \; \exists \alpha' :   \vert \vert \, 
\cdot \,\vert \vert_\alpha$ is Hilbert-Schmidt with respect to 
$ \vert \vert \, \cdot \, \vert \vert_{\alpha'} $) \cite{GV64}.\\ 
\\ 
The case we are dealing with is that of a Hilbert Space ${\cal H}$
(with inner product $\langle \, , \rangle_{\cal H}$) and, for
$ a > 0$, where we consider the characteristic function
\begin{equation}
 \chi_{a,_G} (\xi) = \exp \left\{-{1 \over 2a} \langle
G \xi , \xi  \rangle_{\cal H} \right\}, 
\end{equation}
where $G$ is a positive bounded operator on $\cal H$, corresponding to
the infinite dimensional gaussian measure (with covariance $G$)
formally written
\begin{equation}
 d \mu_{a,_G} (\phi) `` = "  {1 \over Z_{a,_G}} \, \exp
\left\{- { a \over 2 }  \langle G^{-1} \phi , \phi
\rangle_{\cal H} \right\} [{\cal D} \phi ], 
\end{equation}
(where $ Z_{a,_G}$ is a constant such that 
$\mu_{a,_G}({\cal H})=1$) the support of which lies in a
Hilbert-Schmidt extension of $\cal H$. All this can be sumarized in
the single equation
\begin{equation}
 \chi_{a,_G} (\xi) = \int_{\cal H} \; \exp \left\{i
\langle \xi,\phi  \rangle_{\cal H} \right\} d \mu_{a,_G} (\phi),
\end{equation}
where the left hand side involves $G$ (see $(15)$)
and the right hand side involves $G^{-1}$ (see $(16)$). This
generalizes the very well known relation
\begin{equation}
 \exp \left\{-{1 \over 2 } \langle A^{-1}\vec{x},\vec{x}  \rangle
\right\} =  {{k} {(\det A )}^{-{1 \over 2}} }  \int_{\R^n} 
\exp \left\{i \langle \vec{x} , \vec{y} \rangle - { 1 \over 2} \langle
A \vec{y} , \vec{y}\rangle \right\} d \vec{y},
\end{equation}
where $ \vec{x}, \vec{y} \in \R^n$, $k$ is a constant, $A$ denotes a
positive matrix and
$\langle \, ,\rangle$ denotes  the inner product in this space.
Equation $(17)$ defines the function 
$\chi_{a,_G}$ as the {\it Fourier Transform} of the gaussian measure
$\mu_{a,_G}$, which we will denote by $\widehat{\mu}_{a,_G}$. 
\subsection{Gaussian Measures and Duality}
Consider the acyclic complex $(5)$
and Hodge decomposition $(6)$ on the space of $k$-forms.
We take $ {\cal H}_k= L^2 ( \Omega^k)$, where the closure is
taken with  respect to the $L^2$-hermitian product $\langle \,
, \rangle$ defined by the Riemannian metric on $M$ and the inner
product structure of $E(\rho)$, and we consider the decomposition 
${\cal H}_k \cong {\cal H}_k' \oplus {\cal H}_k''$ induced by $(6)$. 
For $a, b > 0$
consider the gaussian measures ${\mu}_a $ on $\Omega^k$ and $\mu_b$ on
$\Omega_k^{\prime}$ defined by the characteristic functions 
\begin{equation}
 {\widehat \mu_a (\alpha_k)} = \exp \left\{- {a \over
2}\langle \alpha_k , \alpha_k \rangle \right\}
\end{equation}
and
\begin{equation}
{\widehat \mu_b' (\eta_{k}')} = \exp \left\{- {b \over
2}\langle \eta_{k}' , \eta_{k}' \rangle \right\}. 
\end{equation}
\begin{prop} Let $a, b > 0$, then
\begin{equation}
\int_{\Omega_k'} d \mu_b' (\eta_k') \widehat{\mu}_a'
(\eta_k') =
\int_{\Omega^k} d \mu_a (\alpha_k) \widehat{\mu}_b' (\alpha_k'). 
\end{equation}
\end{prop}
{\bf Proof.} By $(27)$, $ {\widehat \mu}_a (\xi) = \int \;
\exp \left\{i \langle \xi, \phi \rangle_{\cal H} \right\} d \mu_a (\phi)$, so
\begin{eqnarray*}
\int_{\Omega_k'} d \mu_b' (\eta_k') \widehat{\mu}_a' (\eta_k') & = &
\int_{\Omega_k'} d \mu_b' (\eta_k') \int_{\Omega^k} \exp \left\{i \langle
\eta_k' , \alpha_k \rangle \right\} d\mu_a (\alpha_k)
\\ & = &
\int_{\Omega^k} d \mu_a (\alpha_k) \int_{\Omega_k'} \exp \left\{i \langle
\eta_k' , \alpha_k' \rangle \right\} d\mu_b' (\eta_k')
\\ & =&
\int_{\Omega^k} d \mu_a (\alpha_k) \widehat{\mu}_b' (\alpha_k'). 
\end{eqnarray*} 
\endsquare \\ \\
Let us see that equality between the partition functions
corresponding to the action functionals $(13)$ and $(14)$ can be seen
as the {\it heuristic} limit of $(21)$ when $b$ goes to infinity. Let
$\epsilon$ be a positive real number and take $b = {1 \over
\epsilon}$. Then, 
$$\int_{\Omega_k'} d \mu_{1 \over \epsilon}' (\eta_k') \widehat{\mu}_a'
(\eta_k') = \int_{\Omega^k} d \mu_a (\alpha_k) \widehat{\mu}_{1 \over
\epsilon}' (\alpha_k')$$
and taking the limit $\epsilon \rightarrow 0$ (in the sense of
distributions) of the gaussian characteristic function
$\widehat{\mu}_b'$ we find a Dirac delta function forcing $\alpha_k'$ 
to vanish. The corresponding limit of the associated gaussian measure on 
$\Omega_k'$ is {\it heuristically} (proportional to a) Lebesgue measure 
on that space. Thus, if we write the formal expression (all these
calculations are formal, the measures are of course  ill-defined
``Lebesgue measures'' on $L^2$ spaces of forms)
\begin{equation}
\int_{\Omega_k'}  \widehat{\mu}_a' (\eta_k') \; 
[{\cal D}\eta_k']\;
``=" \int_{\Omega^k} d \mu_a (\alpha_k) \; \delta[\alpha_k' = 0] , 
\end{equation}
we find, by using of the formal relation $(16)$,
\begin{eqnarray*}
\int_{\Omega_k'}
\exp \left\{-{a \over 2}\langle \eta_k',\eta_k'\rangle \right\}
[{\cal D}\eta_k'] 
\; &``="& \int_{\Omega^k} 
\exp \left\{ -{1 \over 2a} \langle \alpha_k , \alpha_k \rangle \right\} 
\delta[\alpha_k' = 0][{\cal D}\alpha_k] \\
&``="& \int_{\Omega^{\prime \prime}_k}  \exp \left\{-{1 \over 2a}
\langle \alpha_k^{\prime \prime} , \alpha_k^{\prime \prime} \rangle 
\right\} [{\cal D}\alpha_k^{\prime \prime}].
\end{eqnarray*}  
Now let us do the change of variables defined by the
maps $(11)$ and $(12)$, $\eta_k^\prime = d_{k-1}\omega_{k-1}^{\prime
\prime}$ and $\alpha_k^{\prime \prime}=
d_{k}^*\omega_{k+1}^{\prime}$, then we find 
\begin{eqnarray}
{\cal J}_{k-1} \int_{\Omega_{k-1}''} \!\!\!\!\!\! \exp \left\{-{a\over 2}
\widehat{\cal S}(\omega_{k-1}'') \right\} & [{\cal D}\omega_{k-1}'']& \\ 
\nonumber
&``="&
{\cal J}_{k+1} \int_{\Omega_{k+1}'} \!\!\!\!\!\! \exp \left\{-{1 \over 2a}
\widehat{\cal S}^*(\omega_{k+1}') \right\} [{\cal D}\omega_{k+1}'] , 
\end{eqnarray}
where ${\cal J}_{k-1}$ and ${\cal J}_{k+1}$ denotes the associated
jacobian determinants ${\cal J}_{k-1}:= \sqrt{ \det( d_{k-1}^*
d_{k-1})}$ and ${\cal J}_{k+1}:= \sqrt{ \det( d_{k }  d_{k }^*)}$. \\
\\
Finally let us write down the formal calculations usually used to
arrive to relation $(23)$; they involve Fourier Transforms, usual
properties of gaussian integrals and changing the order of integration
\cite{Q98}:
\begin{eqnarray*} \int_{\Omega^\prime_k} \!\!\!
& \exp \left\{-{a\over 2} S_o(\eta^\prime_k) \right\}
& [{\cal D}\eta^\prime_k]  \\  &``="&
\int_{\Omega^\prime_k} [{\cal D}\eta^\prime_k] \int_{\Omega^k}
[{\cal D}\alpha_k] \exp \left\{-{1 \over 2a} S_o(\alpha_k )\right\} 
\exp \left\{i\langle \eta_k^\prime, \alpha_k\rangle \right\} \\
&``="& \int_{\Omega^k}  [{\cal D}\alpha_k] \exp \left\{-{1 \over 2a}
S_o(\alpha_k ) \right\} \int_{\Omega^\prime_k} [{\cal D}\eta^\prime_k]
\exp \left\{i\langle \eta_k^\prime, \alpha_k\rangle \right\} \\
&``="& \int_{\Omega^k}  [{\cal D}\alpha_k] \exp \left\{-{1 \over 2a}
S_o(\alpha_k )\right\} \int_{\Omega^\prime_k} [{\cal D}\eta^\prime_k]
\exp\left\{i\langle \eta_k^\prime, \alpha_k^\prime\rangle\right\} 
\\
&``="& \int_{\Omega^k}  [{\cal D}\alpha_k] \exp \left\{-{1 \over 2a}
S_o(\alpha_k )\right\} \delta \left[ \alpha_k^\prime=0\right]\\
&``="& \int_{\Omega^{\prime \prime}_k}  \exp \left\{-{1 \over 2a}
 S_o(\alpha_k^{\prime \prime} ) \right\} [{\cal D}\alpha_k^{\prime
\prime}],
\end{eqnarray*}
which, after the change of variables defined by the maps $(11)$
and $(12)$, is equivalent to $(23)$.
\\ \\
Let us
make a few comments on this computation which, although very
formal, gives the gist of the dualization  procedure.
\begin{enumerate}
\item {\it Hodge decomposition} in the case of an acyclic complex
splits the space of $k$-antisymmetric tensor fields
$(6)$ and then, through isomorphisms $(11)$ and $(12)$, 
\begin{equation}
\Omega^{k} \cong \Omega_{k-1}'' \oplus \Omega_{k+1}'.
\end{equation}
The $L^2$ scalar product on $\Omega^{k}$ then  gives rise to two
(non degenerate) actions
$\widehat{\cal S}$ and $ \widehat{\cal S}^*$ ($(13)$ and $(14)$),
on $\Omega_{k-1}''$ and $\Omega_{k+1}'$  respectively, which are
related by a {\it Fourier transform}. The non-degeneracy in the
actions comes from the fact we restrict ourselves to
\begin{equation}
\Omega_{k-1}'' \stackrel{d_{k-1}}{\longrightarrow} 
\Omega^{k}  
\stackrel{d_{k}^*}{\longleftarrow}  \Omega_{k+1}'.
\end{equation}
Thus, the field $\omega_{k} \in \Omega_{k}$ splits into 
\begin{equation}
\omega_{k}= d_{k-1} \omega_{k-1}'' \oplus d_{k}^*
\omega_{k+1}' ,
\end{equation}
giving rise to two new ``fields" (gauge potentials)
$\omega_{k-1}'',  \omega_{k+1}'$.

\item In the process of taking the Fourier Transform, the
coefficient of  the quadratic action is inverted ($a \mapsto a^{-1}$),
a fact often observed  in duality and typical for Fourier transforms
of gaussian functions. A strong  coupling can thus be turned into a
weak coupling \cite{D98}.

\item Finally, if we consider {\it Hodge star duality} on
the complex, through the relation
$$ d_k^* \omega_{k+1} = (-1)^{nk+n+1} * d_{n-k-1}* \omega_{k+1},$$ 
we recover the usual ``moral" of duality in antisymmetric
fields \cite{Q98}: a $(k{-}1)$-rank antisymmetric tensor field (the
``gauge potential" $ \omega_{k-1}$) is dual to a
$(n{-}k{-}1)$-rank antisymmetric tensor field ($ \eta_{n-k-1} =
*\omega_{k+1}$) or, in ``brane" language, a $(k{-}2)$(electric)-brane is
dual to a $(n{-}k{-}2)$(magnetic)-brane. In fact observe that
\begin{eqnarray*}
 \langle d_k^* \omega_{k+1}, d_k^* \omega_{k+1}
\rangle & = &
\langle * d_{n-k-1}* \omega_{k+1}, * d_{n-k-1}* \omega_{k+1} \rangle
\\
 & = & *^2 \langle d_{n-k-1}\eta_{n-k-1} , d_{n-k-1}\eta_{n-k-1}
\rangle
\end{eqnarray*}
where $*^2$ denotes a $\pm $ sign depending on $k$ and the dimension of
$M$.
\end{enumerate}
From a physical point of view this formal computation tells us that if
we consider an antisymmetric field theory modelling physical fields by
$k$-forms, given by the action $(7)$ then (in this acyclic case) we
find two possible ``potentials" associated to that field:
$\omega_{k-1}''$ and $\omega_{k+1}'$, the first one for the exterior
differential $d_{k-1}$, the second one for $d_{k}^*$ (see $(26)$).
Writing the partition function of the theory with respect to one or
the other give us ``dual" formulations of the same theory. 

\section{Duality and the Analytic Torsion of the de Rham Complex}
Unlike in the previous section, we now want to take into account local 
symmetries of the type $(3)$ and ``dual ones" obtained replacing $d$ by 
$d^*$. Thus we now consider the degenerate actions $(9)$ and $(10)$ 
computing their corresponding partition functions and we show how from 
this point of view duality leads to a factorization of the analytic 
torsion of the space-time manifold. \\
The analytic torsion of a Riemannian manifold $M$ is a topological
invariant defined by some spectral properties of the Laplacian
operators acting on spaces of differential forms on $M$.
These properties are a consequence of the one to one
correspondence
$\Omega_{k}'' \stackrel{d_{k}}{\rightarrow} \Omega_{k+1}'$ used
previously, and its ``dual"
$\Omega_{k+1}' \stackrel{d_{k}^*}{\rightarrow} \Omega_{k}''$ (both of
them defined in the acyclic case), together with
the Hodge star duality map. In this
section we will study the relation between two dual antisymmetric
tensor field actions, their partition functions and the analytic
torsion of the space-time manifold on which such fields are defined. We
will use zeta-regularization techniques
\cite{G95} and an Ansatz introduced by Schwarz to define the partition
function associated to a degenerate action functional \cite{S79}.
\subsection{Zeta-Regularized Determinants and Analytic Torsion on
Riemannian Manifols}   
Let us take again a closed $n$-dimensional
Riemannian manifold $M$ and the acyclic de Rham complex $(5)$ on it, 
with the Hodge decomposition $(6)$ of the space of
($E(\rho)$-valued) $k$-forms on
$M$. The Laplacian operator on $k$-forms,
$ \Delta_k = d_{k-1}d_{k-1}^* + d_k^* d_k$, is a positive selfadjoint
elliptic operator, and its determinant can be defined by the
zeta-function regularization method as,
\begin{equation}
 \det_\zeta \Delta_k = \exp \left\{ -
\zeta_{\Delta_k}' (0) \right\}, 
\end{equation}
where the zeta-function is
given by
\begin{equation}
\zeta_{\Delta_k}(s) = \sum_{\lambda} {1 \over
{\lambda^s}},
\end{equation}
and the sum is over all the eigenvalues
$\lambda$ of
$\Delta_k$. Indeed, it can be shown using properties of elliptic
operators on closed manifolds that this function is analytic for
$s \in \C$ with $Re(s) >>0$, and extends by analytical continuation to
a meromorphic function on $\C$, regular at $s=0$ (see e.g. \cite{G95}).
\\
\\ 
The {\it analytic torsion} of the Riemannian
manifold $M$ \cite{RS71} (see also \cite{R97} for a review) is defined
in terms of the (regularized) determinant of the Laplacian operators
on $ \Omega^k$, $ 0 \le k \le n$, as
\begin{equation}
T(M):=\exp \left( {1\over 2} \sum_{k=0}^n (-1)^{k} k
\, \log (\det_\zeta \Delta_k) \right).
\end{equation}
Since Hodge star duality $*\Delta_k= \Delta_{n-k} *$ implies that
$\Delta_k$ and $\Delta_{n-k}$ are isospectral, so that
$\zeta_{\Delta_k} (s)= \zeta_{\Delta_{n-k}} (s) $ and hence
$\sum_k (-1)^{k } k \zeta_{\Delta_k}(s)$ $ = {n\over 2} \sum_k (-1)^{k } 
\zeta_{\Delta_k}(s)$ $=0$, $T(M)$ is equal to $1$ if $n$
is even. In the odd case it can be shown to be
independent of the metric. \\ \\
Let us recall some simple observations on the
decomposition of the eigenspace of $\Delta_k$ for a given
eigenvalue $\l$. Let us set
${\cal E}_k(\l):=  {\rm Ker}\  (\Delta_k-\l)$, then Hodge decomposition $(6)$
induces a decomposition of such eigenspaces ${\cal E}_k(\l)= {\cal
E}_k^\prime (\l)\oplus  {\cal E}_k^{\prime \prime} (\l)$, where 
$${\cal E}_k^\prime (\l):= \left\{ \omega_k \in {\cal E}_k(\l), d_k
\omega_k=0\right\} = {\cal E}_k(\l) \cap \Omega_k'$$
and
$${\cal E}_k^{\prime \prime } (\l):= \left\{ \omega_k \in {\cal E}_k(\l),
d_{k -1}^*\omega_k=0 \right\} = {\cal E}_k(\l) \cap \Omega_k''.$$ 
Let $\omega_k \in {\cal E}_k (\l)$, and suppose $\omega_k \in
\Omega_k ''$. Then $d_k \omega_k \in \Omega_{k+1}'$ and
$$\Delta_{k+1} d_k \omega_k = d_kd_k^*d_k \omega_k = d_k \Delta_{k}
\omega_k  = \l d_k \omega_k,$$
so $d_k$ maps ${\cal E}^{\prime \prime}_k(\l)$ bijectively into
${\cal E}^{\prime }_{k+1}(\l)$, giving us a bijective correspondence
between (non-zero) eigenvalues (and their corresponding eigenvectors)
of the operators $ d_k^*d_k$ and $d_{k+1}d_{k+1}^*$. 
\\ \\
The zeta-regularization techniques used to define the determinant of
the Laplacian operators can also be used to define the ``regularized
determinants" of the maps $d_{k-1}d_{k-1}^* $ and $d_k^* d_k $.  From
the decomposition $(6)$ of each space $\Omega^k$, and the nilpotency
of the $d_k^*$ and $ d_k $ operators, it follows that the set of 
 eigenvalues of the Laplacian operator $\Delta_k$ is the union of
the  eigenvalues of $d_{k-1}d_{k-1}^* $ and $d_k^* d_k $. 
Since the eigenvalues of $d_{k-1}d_{k-1}^* $ and $d_{k-1}^*
d_{k-1}$ are the same, the set of  eigenvalues of $\Delta_k$
is the union of the $d_{k-1}^*d_{k-1}$ and $d_k^* d_k $
eigenvalues. So, if we define the zeta-function associated to the
operators $d_{k-1}d_{k-1}^*$ and $d_k^* d_k $ by
\begin{equation}
\zeta_{d_{k-1}d_{k-1}^*} (s):=  \sum_{\l''} {1 \over {
{\l''}^{s}}} 
\end{equation}
and
\begin{equation}
\zeta_{d_{k}^*d_{k}} (s):=  \sum_{\l'} {1 \over
{{\l'}^{s}}}, 
\end{equation}
where $\l''$ and $\l'$ denote (non zero) eigenvalues of
$d_{k-1}d_{k-1}^*$ and $d_{k}^*d_{k}$, respectively, it follows
that, for $0 \le k \le n$,
\begin{equation}
 \zeta_{\Delta_k}(s) = \zeta_{d_{k-1}d_{k-1}^*} (s) +
\zeta_{d_{k}^*d_{k}} (s) . 
\end{equation}
Hence
\begin{equation}
 \zeta_{d_{k}^*d_{k}} (s) = \zeta_{\Delta_k}(s)  -
\zeta_{\Delta_{k-1}}(s)  + \zeta_{\Delta_{k-2}}(s) - \cdots +
(-1)^k \zeta_{\Delta_0}(s) ,
\end{equation}  
and from the properties of the
zeta-function of the Laplacian, it follows that
$\zeta_{d_{k-1}d_{k-1}^*}$ and $\zeta_{d_{k}^*d_{k}} $ are well
defined and analytic for $s \in \C$  with $Re(s) >>0$, and extend by
analytic continuation to meromorphic functions on
$\C$, regular at the origin. Moreover, using the fact that
${\cal E}^{\prime \prime}_k(\l) \cong {\cal E}^{\prime }_{k+1}(\l)$,
we find 
\begin{equation}
\zeta_{d_{k-1}^*d_{k-1}} (s) =
\zeta_{d_{k}d_{k}^*} (s). 
\end{equation}
It is clear now that  we can write 
\begin{eqnarray*} \det_\zeta \Delta_k&= & \exp \left\{ - \zeta_{\Delta_k}'(0)
\right\}\\ &= &  \exp \left\{-\zeta'_{d_{k-1}d_{k-1}^*} (0) -
\zeta'_{d_{k}^*d_{k}} (0) \right\}\\
&=& \det_\zeta(d_{k-1}d_{k-1}^*) \,\det_\zeta(d_{k}^*d_{k}),\\
\end{eqnarray*}
where $ \det_\zeta(d_{k-1}d_{k-1}^*) $ and
$\det_\zeta(d_{k}^*d_{k})$ are defined in a similar way as the zeta
determinant of Laplacian operators (equation $(27)$), using $(30)$ and
$(31)$, respectively.

\subsection{The Partition Function of a Degenerate Action Functional
(Following Schwarz)}
Let us come back to the action
${\cal S}_o(\omega_k) = \langle \omega_k , \omega_k
\rangle$
on $ \Omega^k$ and its decomposition $(8)$,  
$ S_o(\omega_k)  = {\cal S}(\omega_{k-1}) \oplus {\cal
S}^*(\omega_{k+1})$, induced by Hodge decomposition of $\Omega^k$,
into the degenerate action functionals ($(9)$ and $(10)$),
$$ {\cal S}(\omega_{k-1}) = \langle d_{k-1}^*d_{k-1} \omega_{k-1} ,
\, \omega_{k-1} \rangle $$
and
$$ {\cal S}^*(\omega_{k+1}) = \langle d_{k}  d_{k}^*\omega_{k+1}, \,
\omega_{k+1}\rangle, \;\;\;\;\;\;\; $$
on $ \Omega^{k-1}= \Omega_{k-1}' \oplus \Omega_{k-1}''
\; (= {\rm Ker}\ (d_{k-1}^*d_{k-1}) \oplus {\rm Ker}\ 
(d_{k-1}^*d_{k-1})^{\perp}) $ and
$\Omega^{k+1}= \Omega_{k+1}' \oplus \Omega_{k+1}'' \; (= 
{\rm Ker}\ (d_{k}  d_{k}^*) \oplus {\rm Ker}\ (d_{k}  d_{k}^*)^{\perp})$,
respectively. In section $2$ we were dealing with non degenerate
actions $ \widehat{\cal S} $ and $ \widehat{\cal S}^*$, since we had
restricted ourselves to $ \Omega_k''$ and $\Omega_{k+1}'$. As we
pointed out there, the degeneracy leads to some formal volume of an
infinite dimensional space, and the non degeneracy condition restrict
us to look at only a part of the complex $(5)$, namely $(25)$. Schwarz
suggested an Ansatz, inspired from the well known Faddeev-Popov
procedure, to ``compute" this volume and give a meaning to the
partition function of a degenerate action functional
\cite{S79}. Following Schwarz's method, provided we can associate a chain of
vector spaces and maps, called {\it resolvent}, to the degenerate
action, then the partition function associated to that action can be
defined in terms of (regularized) determinants of the maps appearing
in the resolvent. In our particular case, as we will see, this means
to consider the whole complex $(5)$ and not only a part of it, as we
done in the non degenerate case.\\ \\
In the case of
${\cal S}(\omega_{k-1})$ the resolvent is the given by (for details
about the definition of the resolvent associated to a degenerate
functional see e.g. \cite{S79} or \cite{BT91})
\begin{equation}
\begin{array}{ccccccccccccc}
 0  \!\!\! & \! 
\stackrel{}{\longrightarrow} \!\!\! & \! \Omega^0 \!\!\! & \!
\stackrel{d_0}{\longrightarrow} \!\!\! &  
\cdots \!\!\! & \! &\stackrel{d_{k-3}}{\longrightarrow}& \!
\Omega^{k-2} \!\!\! & \! \stackrel{d_{k-2}}{\longrightarrow} \!\!\! &
\! \Omega_{k-1}' \!\!\! & \! 
\stackrel{d_{k-1}^*d_{k-1}}{\longrightarrow}
\!\!\! & \! 0, \!\!\!&
\end{array}
\end{equation}
and we define the partition function associated to that action (and
resolvent) as
\begin{equation}
 Z({\cal S})=
\det_\zeta (d_{k-1}^*d_{k-1})^{\frac{-1}{2}}\prod_{j=1}^{k-1}
\vert  \det_\zeta (d_{k-j-1}) \vert ^{(-1)^{j+1}}. 
\end{equation}
Notice that when $k=1$, $Z(S)$ gives back the usual Ansatz to compute 
the partition function (compares with $(4)$ and $(18)$)
$$ Z(S) ``=" \int_{\Omega^0} \exp \{ - {1 \over 2} \langle d_0 f , 
d_0 f \rangle\} [{\cal D} f ]``=" (\det \Delta_0)^{-{1 \over 2}}.$$
In the same way,
taking the resolvent associated to  ${\cal S}^*(\omega_{k+1})$,
\begin{equation}
\begin{array}{ccccccccccccc}
 0  \!\!\! & \! 
\stackrel{}{\longrightarrow} \!\!\! & \! \Omega^n \!\!\! & \!
\stackrel{d_{n-1}^*}{\longrightarrow} \!\!\! & \! 
\cdots \! & \!\stackrel{d_{k+2}^*}{\longrightarrow} \!\!\! & \!
\Omega^{k+2} \!\!\! & \! \stackrel{d_{k+1}^*}{\longrightarrow} \!\!\!
& \! \Omega_{k+1}''
\!\!\! & \! \stackrel{d_kd_k^*}{\longrightarrow} \!\!\! &  \;\;\;  0,
\end{array} 
\end{equation}
we define the associated ``dual" partition function by
\begin{equation}
Z({\cal S}^*)=\det_\zeta (d_{k}d_{k}^*)^{\frac{-1}{2}}
\prod_{j=1}^{n-k-1} \vert \det_\zeta
(d_{k+j}^*) \vert^{(-1)^{j+1}}. 
\end{equation}
Here, $ \det_\zeta(d_{k})$ and $\det_\zeta(d_{k}^*) $ are defined
by
$$ \vert \det_\zeta(d_{k}) \vert  := \sqrt{\det_\zeta(d_{k}^*d_{k})},$$
$$  \vert \det_\zeta(d_{k}^*) \vert := \sqrt{\det_\zeta(d_{k}d_{k}^*)} $$
and, as we remarked in the previous section,
$$\det_\zeta(d_{k}^*d_{k}) = \det_\zeta(d_{k}d_{k}^*) .$$
Therefore, the two ``dual" partition functions are given by
\begin{equation}
 Z({\cal S})= \prod_{j=1}^k [\det_\zeta
(d_{k-j}^*d_{k-j})]^{\frac{(-1)^j}{2}} 
\end{equation}
and
\begin{equation}
Z({\cal S}^*)= \prod_{j=0}^{n-k-1} [\det_\zeta
(d_{k+j}  d_{k+j}^*)]^{\frac{(-1)^{j+1}}{2}}.
\end{equation}

\subsection{Analytic Torsion on Riemannian Manifolds and Duality}
The relation between the analytic torsion of the manifold $M$ and the
partition function of an antisymmetric field theory defined on it, by
the method discussed above, was pointed out by Schwarz
(\cite{S79}, see also \cite{ST84}), studying
quantization of antisymmetric tensor field theories defined by the
degenerate action $(1)$ on $\Omega^k$. It has been
also used in the context of {\it Topological Quantum Field Theories}
\cite{W89}\cite{BT91}\cite{AS95}. Schwarz shows that the Hodge star
duality map and $(34)$ imply a factorization of the
analytic torsion $T(M)$ in terms of the two partition functions,
corresponding to the actions ${\cal S}(\omega_{k-1})$ and ${\cal
S}(\omega_{n-k-1})$.  
\\ \\
In this section we want to relate the
partition functions $Z({\cal S})$ and $Z({\cal S}^*)$ ($(36)$
and $(38)$), corresponding to the two ``dual actions" $ \langle
d_{k-1} \omega_{k-1},d_{k-1}\omega_{k-1} \rangle $ and $ \langle
d_{k}^* \omega_{k+1}, d_{k}^* \omega_{k+1}\rangle$, with the analytic
torsion of the manifold $M$ on which these antisymmetric field theories
are formulated. Such a relation is clear if we look at the splitting in
the de Rham complex $(5)$ induced by the two resolvents, $(35)$ and
$(37)$, associated with the partition functions of the dual theories
defined by the given actions, namely
\begin{equation}
\begin{array}{cccccccccccccccccccccc}
 0  \!\!\! & \! 
\stackrel{}{\longrightarrow} \!\!\! & \! \Omega^0 \!\!\! & \!
\stackrel{d_0}{\longrightarrow} \!\!\! &  \! \cdots \!\!\! & \!
\stackrel{d_{k-2}}{\longrightarrow} \!\!\! &
\! \Omega^{k-1} \!\!\! & \! \stackrel{d_{k-1}}{\longrightarrow}
\!\!\! & \!\! \Omega^k \!\!\! & \! \stackrel{d_k^*}{\longleftarrow}
\!\!\! & \! \Omega^{k+1} \!\!\! & \!
\stackrel{d_{k+1}^*}{\longleftarrow} \!\!\! & \! 
\cdots \!\!\!  & \!\stackrel{d_{n-1}^*}{\longleftarrow} \!\!\! & \!
\Omega^n
\!\!\! & \! \stackrel{}{\longleftarrow} \!\!\! & \!  0.
\end{array} 
\end{equation}
Observe that, from $(39)$ and $(40)$, 
\begin{eqnarray*} \log {\frac {Z({\cal S})}{Z({\cal S}^*)}} = & \,& \log \,
[\det_\zeta(d_{0}^* d_{0})]^{\frac{(-1)^{k}}{2}} + \log \,
[\det_\zeta (d_{1}^* d_{1})]^{\frac{(-1)^{k-1}}{2}} + \cdots \\ &+&
\log \, [\det_\zeta (d_{k-1}^* d_{k-1})]^{\frac{-1}{2}}  
- \log \,[\det_\zeta(d_{k} d_{k}^*)]^{\frac{-1}{2}}\\ & -& 
 \log \,[\det_\zeta (d_{k+1}  d_{k+1}^*)]^{\frac{1}{2}} -
 \cdots -\log [\det_\zeta (d_{n-1}  d_{n-1}^*)]^{\frac{(-1)^{n-k}}{2}},
\end{eqnarray*}
and from $(27)$ and definition $(29)$
$$ T(M) = \exp\left( {1\over 2} \sum_{k=0}^{n-1} (-1)^{k}
\zeta'_{\Delta_k} (0) \right) $$
then,
\begin{eqnarray*}
\log T(M) & = & {1\over 2} \sum_{k=0}^{n-1} (-1)^{k}
\zeta'_{d_k^*d_k}(0) \\
 &=&  \sum_{k=0}^{n-1}  \log (\det_\zeta d_k^*d_k)^{{(-1)^{k}\over 2}}
\end{eqnarray*}
so,
\begin{equation}
{Z({\cal S})}{Z({\cal S}^*)}^{-1} = T(M)^{(-1)^{n-k-1}}.
\end{equation} 
Thus, we can say that two dual actions yield a
factorization of the analytic torsion of the space-time manifold
(coming from the splitting $(41)$) in terms of their corresponding
partition functions. Hence in even dimensions we get the expected 
identification of the partition function with its dual. Note that 
the analytic torsion is a topological
invariant of $M$, but there is no reason for $Z({\cal S}) $ and
$Z({\cal S}^*)$ to have this property.
\\
\\
{\bf Acknowlegments}
This is the written version of two talks given in the Summer School
on {\it Geometrical Methods in Quantum Field Theory}, Villa de Leyva 
(Colombia), July 1999, and at the CIRM colloquium {\it Families of 
Operators and their Geometry}, Marseille (France), June 2000. The author wishes
to thank S. Paycha, T. Wurzbacher, D. Adams and S. Rosenberg for many
helpful discussions. The author is also indebted to the referee for
helpful comments and suggestions.

\end{document}